\documentclass{aastex}
\usepackage{graphicx}

\usepackage{emulateapj5}
\makeatletter

\newenvironment{inlinefigure}{%
\def\@captype{figure}%
\noindent\begin{minipage}{0.999\linewidth}\begin{center}}
{\end{center}\end{minipage}\smallskip} \makeatother

\def\et{{\it et\thinspace al.}}    

\def\msun{{\rm\,M_\odot}}

\def\vol#1  {{{#1}{\rm,}\ }}
\def\lya{{\rm Ly}\alpha}
\def\etal{et al.\ }

\newcount\refno
\newcount\rfno
\def\eq{$^{\the\refno\ }$\advance\refno by 1}
\def\ad{\advance\rfno by 1}

\def\clock{\count0=\time \divide\count0 by 60
     \count1=\count0 \multiply\count1 by -60 \advance\count1 by \time
     \number\count0:\ifnum\count1<10{0\number\count1}\else\number\count1\fi}

\def\myputfigure#1#2#3#4#5%
{\hskip0.03\textwidth\vskip#5pt
\makebox[0pt]{\hskip#2in
\includegraphics[width=#3\textwidth]{#1}}\vskip#4pt\hfill}

\def\Gcm2{\rm G~cm^2}

\def\beq{\begin{equation}}
\def\eeq{\end{equation}}
\def\bea{\begin{eqnarray}}
\def\eea{\end{eqnarray}}

\def \date         {\ifcase\month \message{zero} \or
                    January \or February \or March \or April \or May \or June
                    \or July \or
                    August \or September \or October \or November \or
                    December \fi
                    \space\number\day, \number\year}

\def\pbh{_{{\rm PBH}}}
\def\cdm{_{{\rm CDM}}}
\def\lya{Ly-$\alpha$}
\begin{document}

\title{Primordial Black Holes as Dark Matter: The Power Spectrum and Evaporation
of Early Structures}
\author{N. Afshordi\altaffilmark{1}, P. McDonald,\altaffilmark{2} and D. N. Spergel\altaffilmark{3}
 } \altaffiltext{1}{Princeton University Observatory, Princeton
University, Princeton, NJ 08544;afshordi@astro.princeton.edu}
\altaffiltext{2}{Department of Physics, Princeton University,
Princeton, NJ 08544;pmcdonal@feynman.princeton.edu}
\altaffiltext{3}{Princeton University Observatory, Princeton
University, Princeton, NJ 08544;dns@astro.princeton.edu}

\accepted{ }

\begin{abstract}
   We consider the possibility that massive primordial black holes are the dominant
   form of dark matter. Black hole formation generates entropy
   fluctuations that adds a Poisson noise to the matter power spectrum.
   We use \lya ~forest observations to constrain this Poisson term
   in matter power spectrum, then we constrain the mass of black
   holes to be less than $\sim {\rm few} \times 10^4 \msun$.
   We also find that structures with less than $\sim 10^3$ primordial black holes
   evaporate by now.

\end{abstract}


\section{Introduction}
The nature of Cold Dark Matter (CDM) is one of the most important
and profound
  mysteries in modern cosmology. One of the clues towards solving this
  mystery is the fact that Big Bag Nucleosynthesis (BBN) paradigm
  (e.g., Burles \etal 2001) predicts a much lower density for the baryonic matter
  (by a factor of $5-10$) than almost all the measurements of the CDM density in
  the last decade (see Bahcall \etal~1999 for a review).
  Black holes are an alternative to the most popular candidate for the dark
  matter: massive, non-baryonic, elementary particles (see
  Khalil \& Munoz 2002 for a recent review). If Primordial Black
  Holes(PBHs), formed in the early universe, they could become the dominant
  form of dark matter.

  There are variety of mechanisms proposed for PBH formation: during inflationary reheating
  (Garcia-Bellido \& Linde 1998), from a blue spectrum of primordial density
  fluctuations (Carr \& Lidsey 1993) or during a phase transition in the early
  universe (e.g., Rubin \etal 2000). The typical mass of PBHs at the formation time can be
  as large the mass contained in the Hubble volume, $M_H$, down to around
  $10^{-4}
 M_H$ (Hawke \& Stewart 2002).
  Carr \& Hawking (1974) show that there is no physical self-similar growth
  mode for the PBHs in a radiation gas, and so there will be no
  significant accretion after PBH formation.
  However, Bean \& Magueijo (2002) speculate that the accretion of quintessence (i.e., a scalar field)
  into the PBHs could lead to the scaling of the PBH mass with
  $M_H$.

  In this letter, we consider PBHs with $ 10 \, \msun < M_{\pbh} < \, 10^6 \msun$ as the constituents of
  CDM. The dynamical constraints on such objects has been thoroughly reviewed and discussed by
  Carr \& Sakellariadou (1999). They show that the strongest upper
  limit on the mass of such compact objects in the galactic halo
  is $\sim 10^4 \msun$ which comes from two independent methods:(1) the rate of globular cluster disruptions
  and,(2) the mass of the galactic nucleus, accumulated due to dynamical friction. However, the globular
  cluster disruption limit depends on
  the completely unknown initial number of globular clusters, while the galactic nucleus limit
  ignores black hole ejection through gravitational
  slingshot which may prevent mass accumulation at the center (Xu \& Ostriker 1994). Therefore we
  will take a more conservative
  value of $10^6 \msun$ as our upper limit. On the lower end, the non-detection of long duration lensing events by
  the MACHO microlensing experiment excludes the mass range $0.3-30 \msun$ for the halo
 objects (Alcock \etal 2001).
 This letter studies the impact of these PBHs on the large scale
 structure of the universe, and specifically the signatures of discreteness
 in CDM. In Sec. 2 we study the impact of the Poisson
 noise due to the discreteness of PBHs on the linear power
 spectrum.
  Sec. 3 compares the numerical simulations with this linear power spectrum with
  the observational data of the \lya ~forest power spectrum. This puts an upper limit on the mass
  of PBHs. In Sec. 4 we discuss the effect of evaporation of early
structures, which puts a lower cutoff on the mass function of the
PBH haloes. Finally, Sec. 5 concludes the paper.

\section{The Linear Spectrum}

In the following analysis we assume that PBHs form as the result
of a phase transition at some temperature $T_c$, in the radiation
dominated era. Furthermore, we will assume that there is no
substantial accretion following their formation era and all of
their masses are around $\sim M_{\pbh}$. This assumption can be
generalized to an extended accretion scenario where $T_c$ marks
the end of accretion period (Magueijo and Bean 2002). The black
hole density follows \beq
\rho_{\pbh}=M_{\pbh}n_{0,\pbh}(1+\delta_p)\left(\frac{T}{T_c}\right)^3.
\eeq where $n_{0,\pbh}$ is the number density of PBHs, $\delta_p$
is the fluctuation is the number of PBHs due to Poisson noise, and
the $T/T_c$ factor describes the dilution in PBH density due to
cosmic expansion.
 Since the PBHs are not correlated on acausal distances, one expects that on
scales larger than the Hubble radius \beq P_p=<|\delta_p(k)|^2>=
n_{0,\pbh}^{-1}. \eeq

 For linear perturbations, on large scales, Eq.(1) leads to
\beq \delta_{\pbh} = \delta_p+\frac{3}{4}\delta_{r}, \eeq
 where the $\delta_{\pbh}$,$\delta_p$ and $\delta_r$ are the
 relative overdensities of PBHs, Poisson fluctuations and
 radiation, respectively. Since $\delta_p$ in Eq.(1)is
 observable {\it and} constant, one would conclude that the quantity
 \beq
S\equiv \delta_{\pbh} - \frac{3}{4}\delta_{r} = \delta_p
 \eeq
 is gauge-invariant and conserved. Indeed this is the entropy
 per PBH, which should remain constant as long as the universe
 expands adiabatically (e.g. see Mukhanov \etal 1992). The associated perturbations,
 generated in this way are isocurvature(or entropy) perturbations,
 as the curvature at large scales is not (immediately) affected by
 the formation of compact objects at small scale.

As we are assuming that PBHs are the present day Cold Dark Matter
(CDM), the overdensity of CDM is given by \beq
\delta_{\cdm}(k)=T_{ad}(k)\delta_{i,ad}(k)+T_{iso}(k) S(k), \eeq
 where $T_{ad}(k)$ and $T_{iso}(k)$ are the transfer functions for adiabatic
 and isocurvature perturbations respectively. For the following analysis we will
 use the analytical fits quoted in Bardeen \et ~1986 to the transfer functions. Eq. (5) leads to the following power
 spectrum
 \beq P_{\cdm}(k)=T^2_{ad}(k)P_{i,ad}(k)+T^2_{iso}(k)P_p.\eeq
 In this expression,$P_{i,ad}(k) = A\,
k^n$ with $n\simeq 1$ is the adiabatic power spectrum which is
produced through inflation (or an alternative method of generating
scale-invariant adiabatic perturbations), while $P_p$ is given in
Eq.(2).

  One can easily see that the isocurvature term on the RHS of
  Eq.(2) contributes a constant to the power spectrum as both
  $P_p$ and
\beq T_{iso}(k) = \frac{3}{2}(1+z_{eq}) \, {\rm for} \, k\gg
a_{eq} H_{eq} \eeq
  are independent of $k$ (e.g. Peacock 1998). Note that this
  is the simple linear growth due to gravitational clustering
  which is the same for adiabatic fluctuation. Since the power
  spectrum of adiabatic fluctuations decays as $k^{-3}$ at small
  scales, one expects to see the signature of this Poisson noise
  at large $k$'s. Combining Eqs. (2),(6) and (7) gives the power offset
  \bea
\Delta P_{\cdm} &\simeq& \frac{9 M_{\pbh} (1 + z_{eq})^2}{4
\rho_{\cdm}}\nonumber\\ &=& 4.63 \left( \frac{M_{\pbh}}{10^3
M_{\odot}} \right) (\Omega_{\cdm} h^5) (h^{-1} {\rm Mpc})^3 \eea
which is also a lower bound on the matter linear power spectrum.

\begin{inlinefigure}
\centerline{\includegraphics[width=0.95\linewidth]{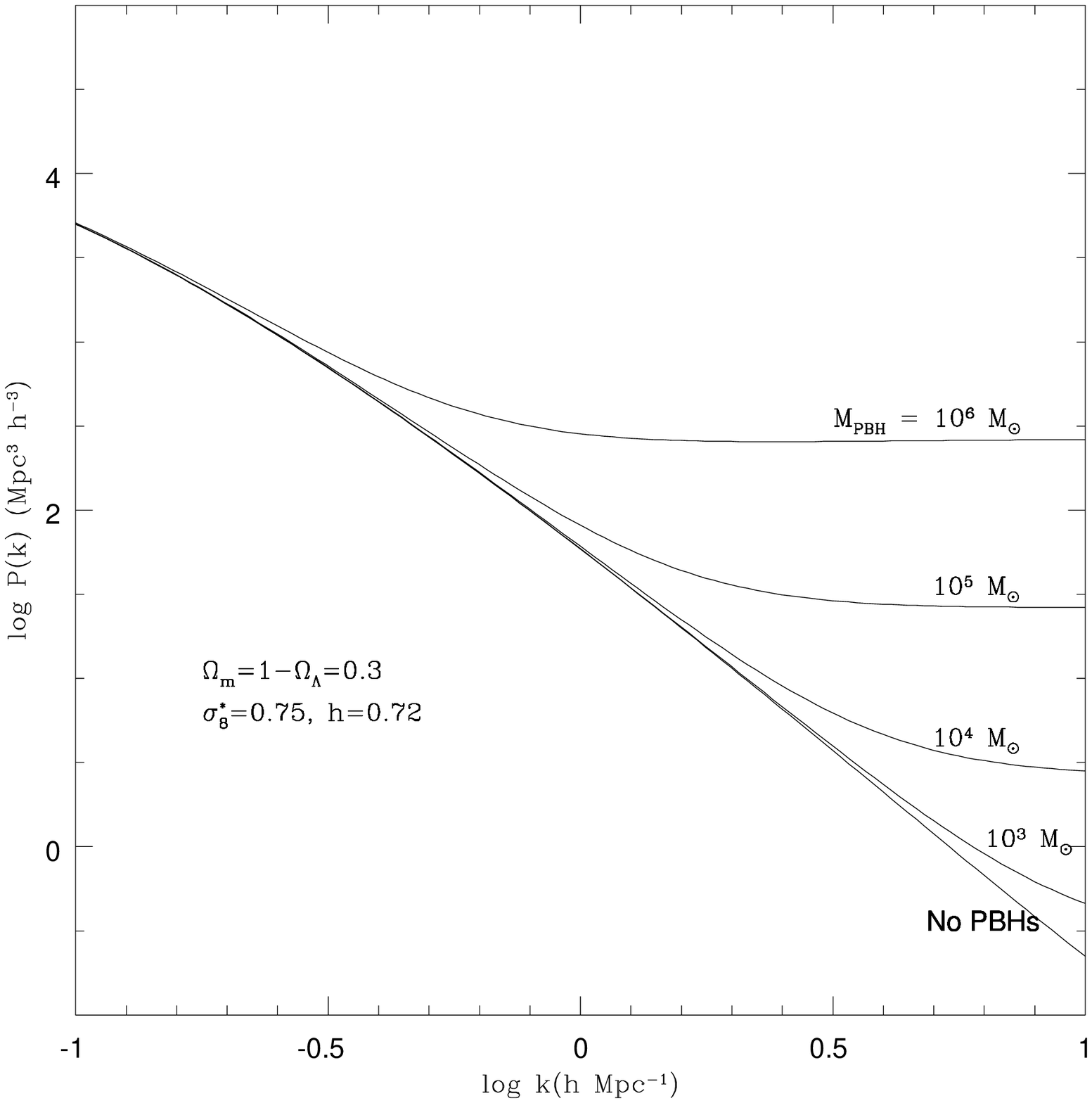}}
\caption{Linear power spectrum for different masses of the PBHs.
$\sigma^*_8$ is $\sigma_8$ for the model without the PBHs and the
amplitude of the (initially) adiabatic modes is the same for all
models.} \label{fig1}
\end{inlinefigure}
 Fig.(1) shows the linear power spectrum for different
masses of the PBHs. We see the Poisson plateau (Eq. 8) at large
k's which drops with decreasing mass. The impact of this plateau
on the \lya ~ forest power spectrum is discussed in the next
section.

\begin{inlinefigure}
\centerline{\includegraphics[width=0.95\linewidth]{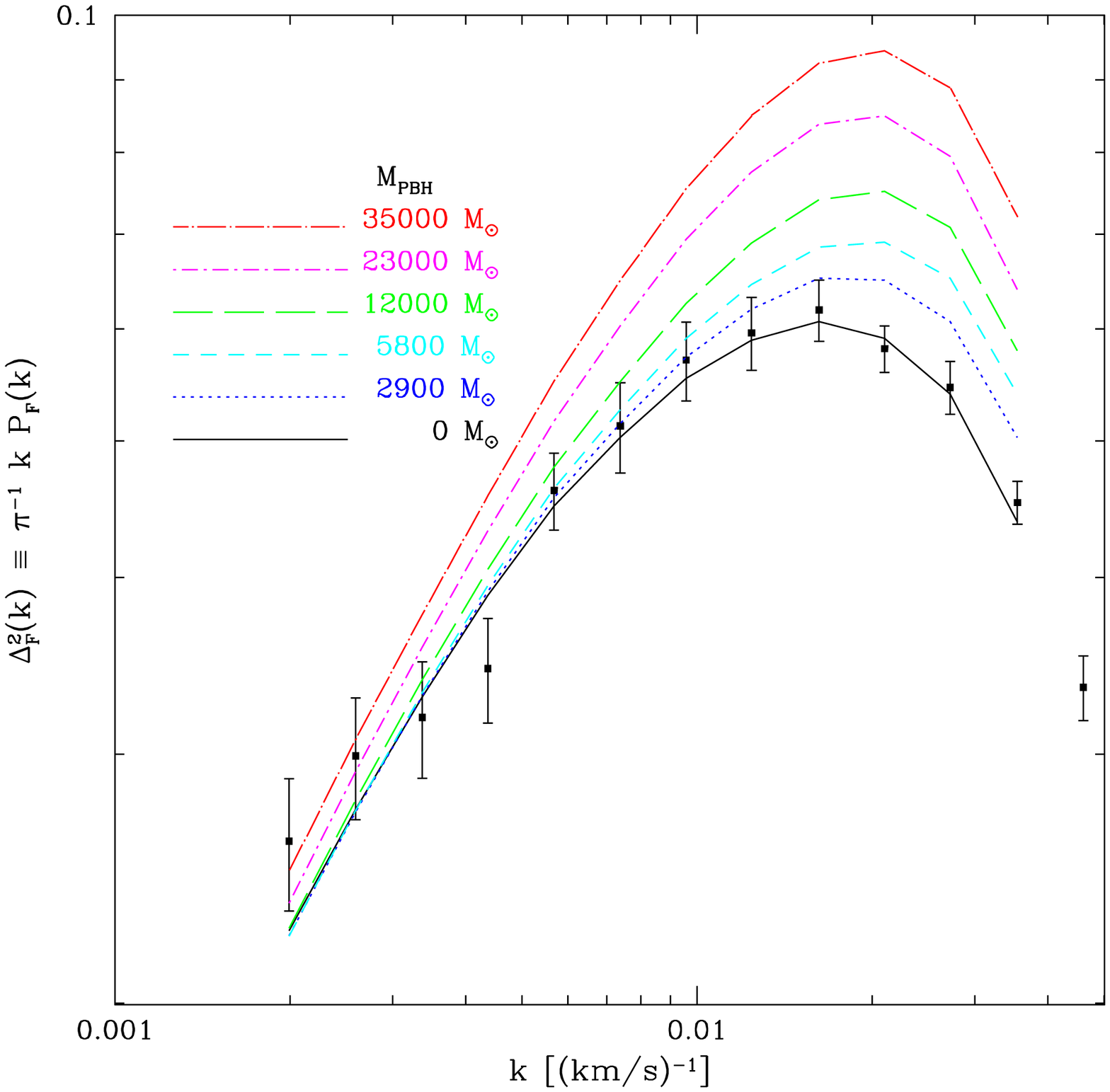}}
\caption{Influence of PBHs on the \lya~ forest flux power
spectrum, $P_F(k)$. The black, solid curve shows our prediction
for $P_F(k)$ in a standard $\Lambda$CDM model (i.e., no PBHs) in
which the amplitude of the linear power spectrum, $\sigma_8^*$,
was adjusted to match the data points from Croft \etal (2002). The
other curves show the predicted $P_F(k)$ when white noise power
due to PBHs with various masses is added. The Ly-$\alpha$ forest
model parameters and $\sigma_8^*$ were not adjusted to find a best
fit for each mass so the disagreement between the PBH models and
the data points does {\it not} indicate that the models are ruled
out. } \label{fig2}
\end{inlinefigure}

\section{Simulations of \lya ~ forest}
The lines in Fig. (2) show the predicted change in the power
spectrum of the Ly-$\alpha$ forest transmitted flux, $P_F(k)$, as
$M_{\pbh}$ is varied.  The points with error bars are $P_F(k)$
measured by Croft et al. (2002) using their fiducial sample
($\bar{z}=2.72$). The predictions were made using the large set of
numerical simulations and the interpolation code described in
McDonald et al. (2003).  We have not yet performed fully
hydrodynamic simulations using PBH power spectra, so our result is
based entirely on hydro-PM simulations (e.g., Gnedin \& Hui 1998;
McDonald, Miralda-Escud\'e, \& Cen 2002; McDonald 2003).  The
curves we show are smooth because the power spectra computed from
the simulations have been compressed into the parameters of an
analytic fitting formula. The background cosmological model used
in Fig. (2) is assumed to be flat with a cosmological constant,
$\Omega_{\cdm}=0.26$, $\Omega_b=0.04$, $h=0.72$, and $n=0.9$ (this
value of $n$ is close to the best fit found by Croft et al. 2002
for our model).  The Ly-$\alpha$ forest model assumed in the
simulation is controlled by 3 parameters:  the mean transmitted
flux fraction in the forest, $\bar{F}$, and the parameters,
$T_{1.4}$ and $\gamma-1$, of a power-law temperature-density
relation for the gas in the IGM,
$T=T_{1.4}(\Delta/1.4)^{\gamma-1}$, where $\Delta$ is the density
of the gas in units of the mean density (see McDonald 2003 for a
demonstration of the effects of these parameters on the flux power
spectrum).  The allowed range of each of these parameters has been
constrained by independent observations. We use the measurement
$\bar{F}=0.746\pm0.018$ from McDonald et al. (2000) and the
measurements $T_{1.4}=20500\pm 2600$ K and $\gamma-1=0.4\pm0.2$
from McDonald et al. (2001). To obtain these values at $z=2.72$,
we interpolated between the redshift bins used by McDonald et al.
(2000, 2001). We subtracted 50\% of the potential continuum
fitting bias they discuss from $\bar{F}$, and add the same number
in quadrature to their error on $\bar{F}$.  We add 2000 K in
quadrature to the error bars on $T_{1.4}$ to help absorb any
systematic errors.  To produce Fig. (2), we fixed these
Ly-$\alpha$ forest parameters to their measured values, and fixed
the normalization of the initially adiabatic component of the
linear power spectrum, $\sigma_8^*$, to the value that gives the
best fit when $M_{\pbh}=0$. It is not surprising to see that the
Ly-$\alpha$ forest power increases dramatically as the white noise
power from the PBHs becomes significant on the observed scales.

Fig. (2) is not sufficient to place constraints on $M_{\pbh}$
because we have not varied any of the other parameters to see if
the predicted power can be adjusted to match the observations.  To
obtain an upper limit on the PBH mass we compute
$\chi^2(M_{\pbh})$, minimizing over the amplitude of the linear
power and the three Ly-$\alpha$ forest parameters, subject to the
observational constraints described above on $\bar{F}$, $T_{1.4}$,
and $\gamma-1$. We follow Croft et al. (2002) in using only
$P_F(k)$ points with $k<0.04$ s/km. Defining an upper limit by
$\chi^2(M_{\pbh})-\chi^2(0)=4$, we find $M_{\pbh}<17000 \msun$.
Fig. (3) shows how this limit is obtained. The temperature-density
relation parameters play no role, but $\bar{F}$ is important. As
$M_{\pbh}$ increases, the best fit value of $\bar{F}$ also
increases until this trend is halted by the external constraint on
$\bar{F}$.  The effect of the increase in $\bar{F}$ is to reduce
$P_F(k)$ (McDonald 2003), counteracting the increase in power due
to $M_{\pbh}$.  Finally, the unconstrained parameter $\sigma_8^*$
also increases with $M_{\pbh}$, further decreasing the power on
small scales while increasing $P_F(k)$ on large scales (see
McDonald 2003; this freedom to adjust the tilt of $P_F(k)$ is
significant -- with $\sigma_8^*$ fixed we find $M_{\pbh}<5800
\msun$).

\begin{inlinefigure}
\centerline{\includegraphics[width=0.95\linewidth]{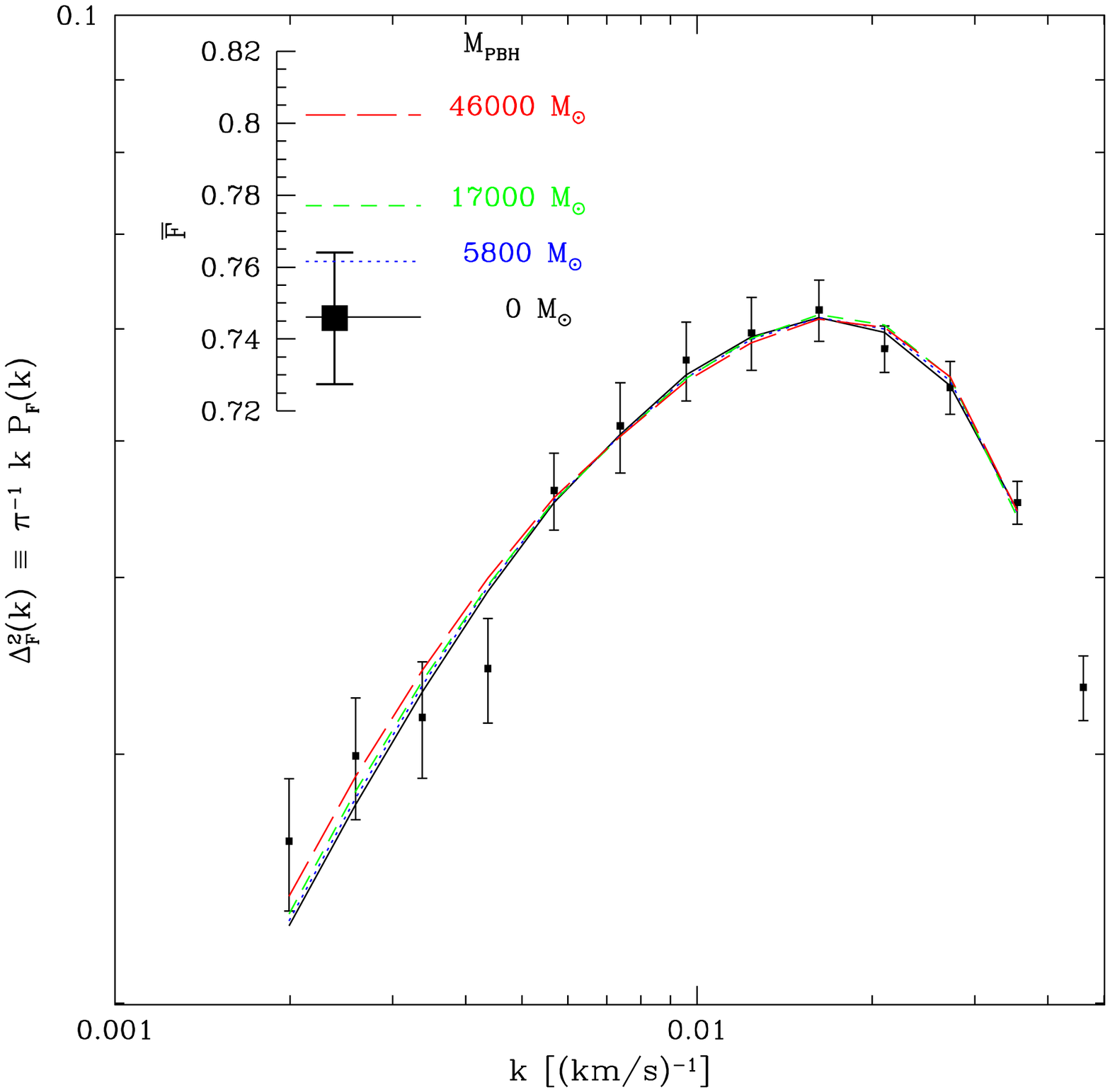}}
\caption{Joint fits to the observed Ly-$\alpha$ forest mean absorption
level, $\bar{F}$, and power spectrum, $P_F(k)$, for several values
of the PBH mass.
Curves show the best predicted $P_F(k)$
for each value of $M_{\pbh}$, determined by a $\chi^2$ minimization
over 4 free parameters:  $\sigma_8^*$, $\bar{F}$, $T_{1.4}$, and $\gamma-1$.
The best fit values of $\bar{F}$ for the different
values of $M_{\pbh}$ are shown by the labeled horizontal
lines.  The large square is the measured value of $\bar{F}$
from McDonald et al. (2000).
}
\label{fig3}
\end{inlinefigure}


Fig. (3) may at first appear unconvincing to the reader unfamiliar with
the Ly-$\alpha$ forest; however, the result is ultimately simple to
understand.  In Fig. (1) we see that the white noise power begins to
dominate on the scales to which the
Ly-$\alpha$ forest is sensitive when $M_{\pbh}\sim 10000 M_\odot$
(note that 1 comoving Mpc/h = 108 km/s at $z=2.72$ in our model).
As $M_{\pbh}$ increases there is simply
too much power on the scale of the Ly-$\alpha$ forest to produce the
observed level of fluctuations.  Increasing $\bar{F}$ can cancel some
of the effect but the size of the increase is limited because $\bar{F}$
is directly observable.

A factor of $\sim 2$ relaxation in the upper bound seems unlikely
but not inconceivable.  For example, if we arbitrarily increase
the error bar on $\bar{F}$ to $\pm 0.03$, the limit we derive is
$M_{\pbh}<41000 M_\odot$.  The limit is $M_{\pbh}<37000 M_\odot$
if we arbitrarily decrease the predicted $P_F(k)$ by 10\% for all
models. The assumed value of $n$ has no effect on the result (we
obtain $M_{\pbh}<18000 M_\odot$ using $n=1$).  Finally, we remind
the reader that the Ly-$\alpha$ forest only constrains the power
spectrum in km/s units at $z=2.72$.  Equation (8) and our assumed
cosmological model were used to compute $M_{\pbh}$.

\section{Early Structures and Relaxation Effects}

  So, apart from the Poisson noise, is there any difference between the gravitational
clustering of the conventional CDM (WIMP particles) and PBHs? The
answer is yes. The collisional relaxation time for a gravitational
system is of the order of the number of particles, times the
dynamical time of the system. Therefore, one expects the
relaxation related effects, e.g. evaporation and core collapse, to
happen faster for systems with smaller number of particles. As the
structures form bottom-up in a hierarchical structure formation
scenario (and even more so in the presence of PBHs as the spectrum
is bluer), and the dynamical time for cosmological haloes is of
the order of the age of the universe, such effects may be
important only for the first structures, that form right after the
recombination era, which have the smallest number of PBHs.

  Let us make a simple estimate of how evaporation of early
  structures sets a lower limit on the mass of virialized objects.
  The evaporation time of an isolated cluster can be estimated
  using
  \beq
  t_{evap} \sim 300 t_{rel} \sim 300\left[\frac{0.14 N}{\ln(0.14 N)}\right] \sqrt{\frac{r^3_h}{GM}},
  \eeq
  (see Binney \& Tremaine 1987), where the subscripts,
  "$evap$" and "$rel$" refer to the
  characteristic times, subsequently associated with the
  evaporation and relaxation of the structure, while $N$
  and $r_h$ are the number of particles and the median radius, respectively.
  To relate $r_h$ to the formation time, for
  simplicity, we consider a truncated singular isothermal sphere which within the spherical collapse model
  (e.g., Afshordi \& Cen 2002) yields
  \beq
  r_h \simeq \frac{2GM}{15 \sigma_v^2} \simeq
  (2GM)^{1/3}\left(\frac{t_{f}}{27\pi}\right)^{2/3},
  \eeq
  where $t_f$ is the formation time of the object.
  Combining this with Eq.(9) gives
  \beq
  t_{evap} \simeq \left[\frac{0.7 N}{\ln(0.14 N)}\right] t_{f}.
  \eeq
  The next assumption is that the
  approximate formation time of the structure is when the variance
  of fluctuations at the mass scale of the structure $\sigma(M)$
  is around the critical overdensity in the spherical collapse
  model, $\delta_c$ (Gunn \& Gott 1972),
 \beq
\sigma(M)|_{_{{\rm form}}} \sim \delta_c \simeq 1.67. \eeq

   Based on the
  calculations of Sec. 2, $\sigma(M)$, which is dominated by the
  Poisson noise at the small scales, is given by
  \beq
  \sigma^2(M) \simeq
  \frac{M_{\pbh}}{M}\left[\frac{3}{2}\left(\frac{1+z_{eq}}{1+z}\right)\right]^2,
  \eeq
  neglecting a late time cosmological constant which is not important at the formation
  time of the early structures. Now, combining Eqs.(11-13) with
  \beq
  \frac{t}{t_f} \simeq \Omega_m^{0.2}
  \left(\frac{1+z_f}{1+z}\right)^{3/2},
  \eeq
   for a flat universe, gives the minimum
  mass for the structure not to evaporate
  \bea
  M_{{\rm min}} &=& N_{{\rm min}} M_{\pbh} \nonumber\\
  &\sim& M_{\pbh} \left[\frac{3(1+z_{eq})}{2\delta_c(1+z)} \right]^{6/7}
  \left[\frac{\Omega^{0.2}_m \ln(0.14 N_{{\rm min}})}{0.7}\right]^{4/7}\nonumber\\
  &\simeq& 3 \times 10^3 \, M_{\pbh} (1+z)^{-1} \left(\frac{\Omega_m
  }{0.3}\right)\left(\frac{h}{0.7}\right)^{12/7},
  \eea
    and consequently, the structures with $M<M_{{\rm min}}$ should
    evaporate by redshift $z$.

    This calculation may be an underestimate of $M_{{\rm min}}$
    since Eq.(9) is valid for an isolated object and ignores tidal effects.
    In general, tidal stripping can decrease $t_{evap}$ significantly (by a factor
    of 5-10). However, accretion, on the other hand, can slow down the net mass loss rate of the
    structure and hence has a reverse effect. In reality, the
    exact lower limit will probably depend on the environment of
    the structures.
\section{Conclusions}

  We have studied the possibility of having CDM in the form of
  PBHs (or any other massive compact object) and its impact on the large scale
  structure of the universe. We see that the simple Poisson noise,
  enhanced by gravitational clustering in the matter dominated
  era, leads to a plateau in the power spectrum at large wave
  numbers (see Fig. 1). Comparison of numerical simulations of \lya ~ forest
  with the current observational data rules out
  the PBH masses larger than $ {\rm few} \times 10^4 \msun$.
  The discrete nature of the PBHs can also lead to the evaporation
  of small (early) structures. A simple estimate shows that this
  puts a lower limit of about $10^3 M_{\pbh}$ on the mass of
  small structures.

\acknowledgments PM thanks Nick Gnedin for the hydro-PM code.
Also, NA and DNS are grateful to J.P.Ostriker for critical and
illuminating discussions. The Ly-$\alpha$ forest simulations were
performed at the National Center for Supercomputing Applications.

\end{document}